\documentclass[twocolumn,psfig,amssymb]{aa}

\usepackage{graphicx,natbib}
\bibpunct{(}{)}{;}{a}{}{,} 
\begin{document}

\renewcommand{\vec}[1]{\boldsymbol{#1}}
\newcommand{\pd}[2]{\frac{\partial{#1}}{\partial {#2}}}

\title{Stability and nonlinear adjustment of vortices in Keplerian flows}

\author{G. Bodo\inst{1} \and A. Tevzadze\inst{2} \and 
G. Chagelishvili\inst{2} \and A. Mignone\inst{1,3}  and P. Rossi \inst{1} \and A. Ferrari\inst{3} 
            }


   \institute{
Osservatorio Astronomico di Torino, Strada dell'Osservatorio
             20, I-10025 Pino Torinese\\
             email: bodo@to.astro.it
\and
 E. Kharadze Georgian National Astrophysical Observatory,, 2a Kazbegi Ave. Tbilisi 0160, Georgia  
\and
Dipartimento di Fisica Generale dell'Universit\`a,
Via Pietro Giuria 1, I - 10125 Torino
                          }

\date{Received ; accepted }

\abstract{} { We investigate the stability, nonlinear development
and equilibrium structure of vortices in a background shearing Keplerian flow } 
{We make use of high-resolution global two-dimensional compressible 
hydrodynamic simulations. We introduce the concept of nonlinear adjustment 
to describe the transition of unbalanced vortical fields to a
long-lived configuration. } 
{ We discuss the conditions under
which vortical perturbations evolve into  long-lived  persistent
structures  and we describe the properties of these equilibrium
vortices. The properties of equilibrium vortices appear to
be independent from the initial conditions and depend only on the
local disk parameters. In particular we find that the ratio of the
vortex size to the local disk scale height increases with the
decrease of the sound speed, reaching values well above the unity.
The process of spiral density wave generation by the vortex,
discussed in our previous work,
appear to maintain its efficiency also at nonlinear amplitudes and
we observe the formation of spiral shocks attached to the vortex.
The shocks may have important consequences on the long term vortex
evolution and possibly on the global disk dynamics.} 
{ Our study strengthens the arguments in favor of anticyclonic vortices as
the candidates for the promotion of planetary formation.
Hydrodynamic shocks that are an intrinsic property of persistent
vortices in compressible Keplerian flows are an important
contributor to the overall balance. These shocks support vortices 
against viscous dissipation by generating local potential
vorticity and should be responsible for the eventual fate of the
persistent anticyclonic vortices. Numerical codes have be able
to resolve shock waves to describe the vortex dynamics correctly. }


\authorrunning{Bodo et al.}
\titlerunning{Stability and nonlinear adjustment of vortices}

\maketitle

\section{Introduction}

Recent advances in the understanding of vortex behavior in
differentially rotating flows are mainly associated with the study
of protoplanetary disk dynamics, since \cite{von44aa} and
\cite{AdaWat95aa}, who suggested that vortices can promote the
formation of planetesimals. Indeed, it has been shown that vortices,
if sustained long enough, lead to particle aggregation in their core
\citep[see e.g.][]{Cha00aa, de-Bar01aa, JohAndBra04aa, KlaBod06aa} and to the
formation of protoplanets.

The vortex scenario for planetary formation encounters an apparent
obstacle: any structure in a Keplerian disk is subject to a strong
shearing  that may eventually lead to its  decay. The only mechanism
for  sustaining a stable vortex in such flows is nonlinearity.
Hence, vortices that may start the process of planetary formation
should exceed a critical threshold in their amplitude. Direct
numerical simulations are therefore an important tool in these
studies.

Previous studies on vortex stability, performed mainly by numerical
simulations \citep[see][]{BarSom95aa, BraChaPro99aa, GodLiv99aa,
GodLiv99ab, DavSheCuz00aa, Dav02aa, KlaBod03aa, Kla04aa,  BarMar05aa, UmurReg04aa}, have
shown that only anticyclonic vortices can be nonlinearly stable in
Keplerian flows and can promote dust trapping and the creation of
protoplanets.  Similar indications come from  hydrodynamical
simulations of accretion disks with initially imposed random
vorticity fields, where again only anticyclonic vortices survive
\citep{SheStoGar06aa, JohGam05aa}.

Although different numerical simulations have already shown that
anticyclonic vortices can survive in compressible Keplerian disks,
most of the stable structures are observed on scales smaller than
the scale height of the disk. This calls for three-dimensional
investigations before solid conclusions can be drawn. On the other
hand, \cite{BarMar05aa} have found indications of a 3D vortex
instability and have shown that only off mid-plane vortices with a very
specific vertical structure are stable in 3D. The preferred location
for dust capture and planetesimals are, however, the high density
regions in the midplane of the disk, where vortices are unstable. A
solution to this problem could be the existence of vortices with a
horizontal extent larger than the disk thickness. Those should in
fact behave as 2D structures therefore being stable and promote mass
accumulation. However, presently, there are doubts on the
possibility of the existence of such vortices
\citep[see][]{JohGam05aa, BarMar05aa}. They are thought to decay to
a smaller size, of the order of the disk scale height, due to the
supersonic velocity circulation.

In the present paper, we study the nonlinear dynamics of relatively
large scale (exceeding the disk half thickness) vortices in
keplerian flows by  global compressible simulations. In this
situation, compressibility is an important factor that cannot be
neglected because the Keplerian profile, setting a strong radial
velocity shear, couples necessarily  vortex and wave mode
perturbation \citep[see][]{BodChaMur05aa}. Our aim is to study in
detail equilibrium structure of nonlinear
long lived vortices. In our study we assume the initial vortical
  perturbation as given and we investigate how it evolves into an
  equilibrium configuration and the properties of this
  configuration. It is not our aim here to study how such large
  amplitude perturbation was formed. Recent studies 
  \citep[see e.g.][]{KlaBod03aa, PetSteJul07aa}  have pointed
  out how baroclinic effects can lead to a growth of vortical
  perturbations, {\bf but that a radial temperature variation is required.
  However, here, for studying the vortex structure, we assume, for
  simplicity, a disk with  initially uniform temperature. The adequacy of this approximation must be tested through future studies}  We will focus our
analysis on the possible existence of 
vortices with horizontal extent larger than the disk scale height.
An important process for the vortex dynamics in compressible flows
is the linear generation of spiral density waves, described in
\cite{BodChaMur05aa}. In the present paper we further investigate
this process by verifying its efficiency during the nonlinear stages
and by analyzing its relevance for the vortex structure. In this
respect, an interesting point is that equilibrium vortex structures
work as a mill that processes the shear flow energy into coherent
wave emission. The nonlinear development of spiral-density waves
leads to the formation of spiral shocks with a steady pattern. These
shocks may increase the stability of anticyclonic vortices by
slowing down their decay and may have also global effect on the
disk.

In \S \ref{sec:setup} we describe the numerical code, setup and
initial conditions used in the simulations. In \S \ref{sec:results}
we present the results of our numerical simulations and describe the
nonlinear adjustment to an equilibrium configuration. We consider
the effect of the  vortex initial amplitude and size on its further
evolution and analyse the timescales of the adjustment process. In
\S \ref{sec:structure} we discuss the stability and structure of the
developed  vortex structures. In \S
\ref{sec:waves} we study the process of wave emission  induced  by
the vortex and the subsequent shock development. A study on the
effects of numerical resolution is given in \S \ref{sec:numer}. The
paper is finally summarized in \S \ref{sec:summary}.

\section{Numerical code and setup}\label{sec:setup}
%
%
%
%

We solve the equations of inviscid compressible gas dynamics in two
dimensional polar coordinates $(r,\phi)$. From the observer's frame,
let $\rho$, $m_r$, $L_\phi$ and $E$ be, respectively, the fluid
density, radial, angular momentum and total energy density; their
conservation is expressed by
\begin{eqnarray}
 \pd{\rho}{t} + \frac{1}{r}\pd{(r\rho v_r)}{r}
              + \frac{1}{r}\pd{(\rho v_\phi)}{\phi} & =& 0  \,, \label{eq:dens}\\
 \pd{(\rho v_r)}{t} + \frac{1}{r}\pd{(r\rho v^2_r)}{r}
                    + \frac{1}{r}\pd{(\rho v_rv_\phi)}{\phi}
                    + \pd{p}{r}& =& \frac{\rho v_\phi^2}{r} + \rho g_r \,,      \\
 \pd{L_\phi}{t} + \frac{1}{r}\pd{(r L_\phi v_r)}{r}
                + \frac{1}{r}\pd{(L_\phi v_\phi + rp)}{\phi}& =& 0         \,,    \\
 \pd{E}{t} + \frac{1}{r}\pd{\left[r(E + p)v_r\right]}{r}
           + \frac{1}{r}\pd{\left[ (E + p)v_\phi\right]}{\phi} &= &m_rg_r\,,
\label{eq:en}
\end{eqnarray}
where $v_r$ and $v_\phi = L_\phi/(\rho r)$ are the radial and
azimuthal velocities, $p$ is the thermal pressure and $g_r = -1/r^2$
is the gravitational acceleration. The total energy $E$ of the fluid is
  expressed as the sum of the kinetic and internal energy via the
  ideal equation of state:
  \begin{equation}
 E = \frac{p}{\Gamma-1} + \rho \frac{v_{r}^{2}  + v_{\phi}^{2}}{2}    
  \end{equation}
where $\Gamma = 5/3$ is assumed. 
Dissipative effects are only those related to the use of a mesh of
finite width
Equations (\ref{eq:dens})--(\ref{eq:en})
are solved using the hydrodynamics module of the PLUTO code
\citep{MigBodMas07aa}. PLUTO evolves the conservative equations in a
Godunov-type fashion by first interpolating volume averages inside
each cell and then solving a Riemann problem at each interface to
compute the fluxes. For the present work, we choose second order
slope-limited interpolation and the two-stage Runge Kutta time
stepping scheme of \cite{GotShu98aa}. An accurate Riemann solver
based on the two-shock approximation \citep{ColWoo84aa} is used to
compute numerical fluxes.

Integration of the equations is made considerably faster by taking
advantage of the FARGO scheme \citep{Mas00aa}, available in the
PLUTO code. This scheme allows larger timesteps than the standard
integration by removing the average azimuthal velocity from the
Courant condition, which severely limits the time step because of
the fast orbital motion at the inner boundary. Besides the increased
computational efficiency, this technique considerably reduces the
amount of numerical diffusion in the solution.

Our computational domain covers the region $0\le\phi\le 2\pi$,
$R_{\rm in}\le r\le R_{\rm out}$, where $R_{\rm in}$ and $R_{\rm
out}$ are the innermost and outermost boundaries in the radial
direction. The grid is constructed by first assigning the number
$N_\phi$ of uniform zones in the $\phi$ direction, whence
$\Delta\phi = 2\pi/N_\phi$. The mesh spacing in the radial direction
is then chosen to yield approximately equal radial and azimuthal
lengths $\Delta r_i = r_i\Delta\phi$ on each ring $i=1,\cdots,N_r$
of the disk. The number of radial zones $N_r$ is determined by the
condition $r_i + \Delta r_i/2 = r_{i+1} - \Delta r_{i+1}/2$ (grid
continuity) at each node, giving a first (non-integer) estimate
\begin{equation}\label{eq:nr}
N^*_r = {\log \left( {R_{\rm out} \over R_{\rm in}} \right) / \log
\left( {N_\phi + \pi \over N_\phi - \pi} \right)} .
\end{equation}
Finally, we obtain $N_r$ by rounding $N^*_r$ up to the nearest
integer which requires  $R_{\rm out}$ to be properly adjusted by
inverting Eq. (\ref{eq:nr}) with $N^*_r$ replaced by $N_r$. Hence,
the radial resolution is set up by the azimuthal spacing and by the
innermost and outermost boundaries.

The initial condition, at $t = 0$, consists of a Keplerian disk
  with uniform density and pressure plus a velocity perturbation
  described in the following subsection.

Boundary conditions are set as follows: at the outermost radius free
outflow is permitted, whereas at the innermost radius variables are
kept constant to their initial value. Periodic boundary conditions
hold in the azimuthal domain.

For our simulations we used four main numerical setups, with  low,
medium, high and very high resolutions (called respectively L, M, H
and VH), described in detail in Table \ref{tab:numsetup}.
\begin{table}
 \caption[]{Resolution, radial domain and number of revolutions
            \textbf{at $r=1$} for the different numerical setups. The
            domain length in the angular direction is $\phi\in[0,2\pi]$.}
 \label{tab:numsetup}
 \centering
 \begin{tabular}{l c c c}
 \hline\hline
       & Resolution           & Domain            &   Time \\
       & ($N_\phi\times N_r$) & $[R_{\rm in},R_{\rm out}]$ & (Revol.) \\
 \hline
 L.......  & $1500\times550$  & $[0.2, 2]$    & $80$ \\
 M.......  & $2000\times622$  & $[0.4, 4]$    & $200$ \\
 H.......  & $4000\times1466$ & $[0.2, 2]$    & $80$ \\
 VH......  & $8000\times1559$ & $[0.5, 1.7]$  & $80$ \\
 \hline
\end{tabular}
\end{table}

\subsection{Initial vortex}\label{sec:init}
%
%

Initial conditions for numerical simulations are composed by the sum
of the Keplerian flow with $V_\phi = r^{-1/2}$ and  a vortex
perturbation, that can have different geometries, size and
amplitude. The initial vortex configuration most widely used in our
simulations is the following:
\begin{equation}\label{eq:vxpert}
v_x^{\prime}(0)= \pm {\epsilon (y-y_0) \over q} \exp
\left[-{(x-x_0)^2 \over a^2 } - {(y-y_0)^2) \over q^2 a^2} \right] ,
\end{equation}
\begin{equation}\label{eq:vypert}
v_y^{\prime}(0)= \mp \epsilon q (x-x_0) \exp \left[-{(x-x_0)^2 \over
a^2 } - {(y-y_0)^2) \over q^2 a^2} \right] ,
\end{equation}
\begin{equation}\label{eq:rhopert}
\rho^\prime(0) = 0 .
\end{equation}
Here $\epsilon$ defines the amplitude of the initial perturbation
and its sign determines the vortex polarity (positive in the case of
anticyclonic vortex). The parameters $a$ and $q$ describe,
respectively, the size (in the radial direction) and aspect ratio of
an elliptic vortex. A circular vortex corresponds to $q=1$, and $q >
1$ refers to a vortex elongated in the azimuthal direction. We
choose the location of the vortex $(x_0,y_0)$ so that it corresponds
to the radial location $r_0=1$. With these choices, our unit of
length is the radius $r_0$ at which the vortex is located and our
unit of time is the inverse of the Keplerian angular velocity
$\Omega_{\mathrm{Kep}} (r_0)$ at the same location. From a physical
point of view, the vortex amplitude $\epsilon$ is related to the
maximum vorticity $\Omega_M$ (at the vortex center), $\Omega_M = 2
\epsilon$

A second type of initial vortex with algebraic distribution of
potential vorticity is also used,
\begin{equation}\label{eq:vxpert2}
 v_x^{\prime}(0)= \pm {\epsilon (y-y_0) \over q} \left( 1 +
 {(x-x_0)^2 \over a^2 } + {(y-y_0)^2) \over q^2 a^2} \right)^{-1} ,
\end{equation}
\begin{equation}\label{eq:vypert2}
 v_y^{\prime}(0)= \mp \epsilon q (x-x_0) \left( 1 + {(x-x_0)^2 \over
 a^2 } + {(y-y_0)^2) \over q^2 a^2} \right)^{-1} .
\end{equation}
This initial vortex is used for the sake of comparison to see how
different initial potential vorticity fields with similar amplitude
and scale but different geometry behave in the nonlinear regime.

\section{Vortex evolution  and nonlinear adjustment}
\label{sec:results}
%
%
%
%

\subsection{The simulations}
%
%

The evolution of the perturbations depends on three main parameters:
the amplitude and size of the perturbation (respectively $\epsilon$
and $a$) and the sound speed in the disk, $c_s$. Our first aim is to
determine the region, in the parameter space described by $\epsilon,
a$ and $c_s$, in which the evolution of the initial perturbation
leads to a stable, long-lived equilibrium configuration. As we shall
see, before reaching this final state the system undergoes, in the
course of several disk revolutions, a transition phase that we call
{\sl nonlinear adjustment}. Our second aim is to provide a detailed
description of the equilibrium vortex configuration. With these
purposes we have performed runs with different values of the three
parameters and different numerical setups, using Eqs.
\ref{eq:vxpert}-\ref{eq:vypert} with $q = 1$ (circular vortices). In
particular, by increasing the value of the sound speed ($c_s= 0.001,
0.01, 0.1$), we have explored the behavior of the vortex changing
$a$ at fixed $\epsilon$ and changing $\epsilon$ at fixed $a$.
Additional calculations have been performed for the purpose of
exploring in more detail particular regions of the parameter space.
For instance, additional values of $c_s$ have been used for a better
understanding of the scaling behaviors of some of the vortex
properties. Moreover, in order to investigate whether the general
behavior is changed by varying the shape and structure of the
initial perturbation, we have performed computations with different
values of the ellipticity parameter $q$ and of the (algebraic)
potential vorticity distribution (see Sect. above). A complete list
of all the simulations and related parameters are described in Table
\ref{tab:numpar}.

\begin{table}
 \caption[]{Table of parameters used for the computations.
From the left to right, the columns give the sound speed ($c_s$),
the vortex initial size ($a$), amplitude ($\epsilon$), ellipticity
parameter $q$, geometry (namely the shape of the initial vorticity
distribution: exp and alg stand for the initial configurations
given by Eqns. \ref{eq:vxpert},\ref{eq:vypert} and
\ref{eq:vxpert2},\ref{eq:vypert2}, respectively) together with its
polarity $Pol$ ($-$ for anticyclonic vortices and $+$ for cyclonic ones).
 The numerical setup
(NS, given in Table \ref{tab:numsetup}) and an identification label
are given in the {\bf rightmost} columns.}
 \label{tab:numpar}
 \begin{tabular}{ccccccc}
 \hline
 $c_s$ & a & $\epsilon$ & $q$ & Geom/Pol & NS & Label \\
\hline
  0.1 & 0.1 & 0.1 & 1 & exp/ - & L & A1 \\
      &     & 0.2 & 1 & exp/ - & L & A2 \\
      &     & 0.3 & 1 & exp/ - & L & A3 \\
      &     & 0.5 & 1 & exp/ - & H & A4 \\
      &     & 0.8 & 1 & exp/ - & L & A5 \\
      & 0.3 & 0.5 & 1 & exp/ - & L & A6 \\
      & 0.4 & 0.5 & 1 & exp/ - & L & A7 \\
\hline
 0.05 & 0.3 & 0.5 & 1 & exp/ - & M & B1 \\
\hline
 0.01 & 0.01 & 0.5  & 1   & exp/     -   & VH & C1 \\
      &      & 1.   & 1   & exp/     -   & H  & C2\\
      & 0.02 & 0.5  & 1   & exp/     -   & H  & C3  \\
      & 0.05 & 0.5  & 1,2 & exp,alg/ -   & VH & C4 \\
      & 0.1  & 0.1  & 1   & exp/ -   & H & C5 \\
      &      & 0.2  & 1   & exp/ -   & H & C6 \\
      &      & 0.22 & 1   & exp/ -   & H & C7 \\
      &      & 0.25 & 1   & exp/ -   & H & C8 \\
      &      & 0.28 & 1   & exp/ -   & H & C9 \\
      &      & 0.3  & 1   & exp/ -   & H & C10 \\
      &      & 0.5  & 1   & exp/ -   & H & C11 \\
      & 0.2  & 0.2  & 1   & exp/ -   & M & C12 \\
      &      & 0.5  & 1   & exp/ -,+ & H & C13 \\
\hline
 0.005 & 0.02  & 0.5 & 1 & exp/ - & H & D1 \\
       & 0.025 & 0.5 & 1 & exp/ - & H & D2 \\
\hline
 0.002 & 0.02 & 0.5 & 1 & exp/ - & H & E1 \\
\hline
 0.001 & 0.01 & 0.5 & 1 & exp/ - & H & E2 \\
       & 0.02 & 0.5 & 1 & exp/ - & H & E3 \\
       & 0.05 & 0.5 & 1 & exp/ - & H & E4 \\
       & 0.1  & 0.5 & 1 & exp/ - & H & E5 \\ \hline
\end{tabular}
\end{table}

\subsection{Nonlinear amplitude thresholds}
%
%

It is well known that vortices with small amplitudes are sheared
away by  the linear deformation induced by the background flow. 
 Nonlinear effect can counteract the shearing deformation 
 only for vortices with
higher amplitude. In this respect, we found two threshold values for
$\epsilon$ defining the character of the evolution of the initially
imposed vortex. The first threshold value between  linear and
nonlinear behaviors is:
\begin{equation}
 \epsilon^* = 0.1 .
\end{equation}
For $\epsilon<\epsilon^*$ the vortex behaves linearly
\citep{BodChaMur05aa}. Its dynamics is characterized by the
geometrical stretching of the initial configuration, due to the
strong radial Keplerian velocity shear. The linear vortex is then
stretched to fill the entire ($2\pi$) azimuthal domain, within the
first 1.5 local disk revolutions. It then experiences further decay
down to the dissipation scales, where it is damped.

\begin{figure*}[thp!]
 \centering
 \includegraphics[width=17.5cm]{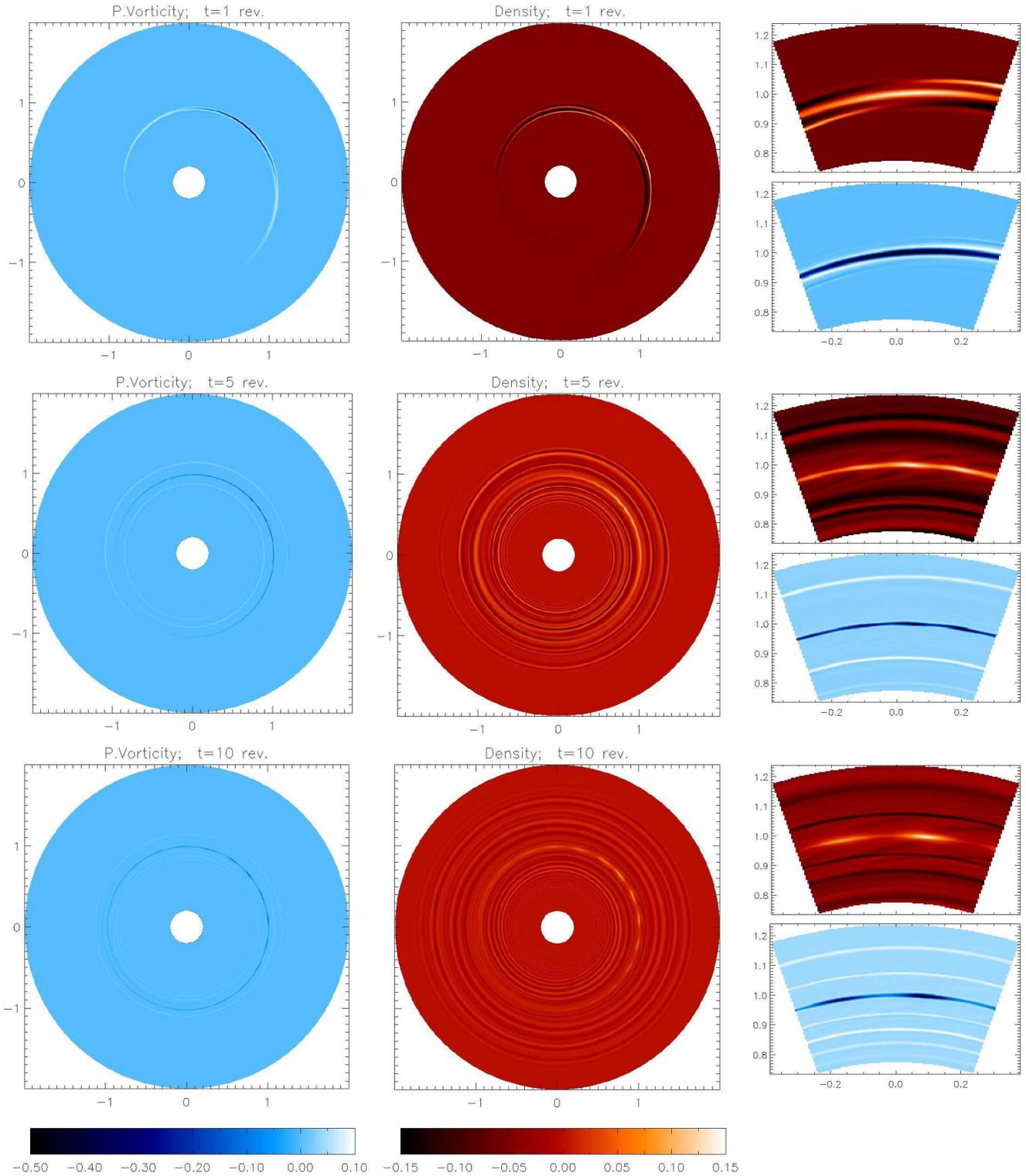}
 \caption{Dynamics of an initially imposed small-scale
 anticyclonic vortex (exponential distribution of vorticity, $q=1$, 
 $a=0.1$) with amplitude exceeding the first nonlinear
 threshold parameter: $\epsilon = 0.2$. The system parameters
 correspond to  case C6 (see Table 2). The first column shows the
 evolution of the perturbed potential vorticity and the second column 
 the evolution of density at times 1, 5 and 10 (in units of the local disk
 revolution period). The third column displays zooms at the
 vortex location at the same times.  The vortical perturbation is
 initially stretched into a narrow vortex sheet which then undergoes local
 nonlinear instabilities producing a chain of small-scales
 anticyclonic vortices that tend to fill the
 entire azimuthal domain.  See electronic version of the
 paper for color figures.  } \label{Fragmentation}
 \end{figure*}

When the vortex amplitude exceeds the first nonlinear threshold, a
two-stage process occurs. First, the vortex is sheared into a narrow
vortex layer, which then undergoes local instabilities. We then
observe the formation of small-scale weak anticyclonic vortices at
different azimuthal locations. Hence, the initial vortex is
transformed into a chain of small scale vortices that tend to spread
and occupy the entire azimuthal domain. The typical character of the
evolution of potential vorticity for this case is shown in Fig.
\ref{Fragmentation}.

\begin{figure*}[thp!]
 \centering
 \includegraphics[width=\hsize]{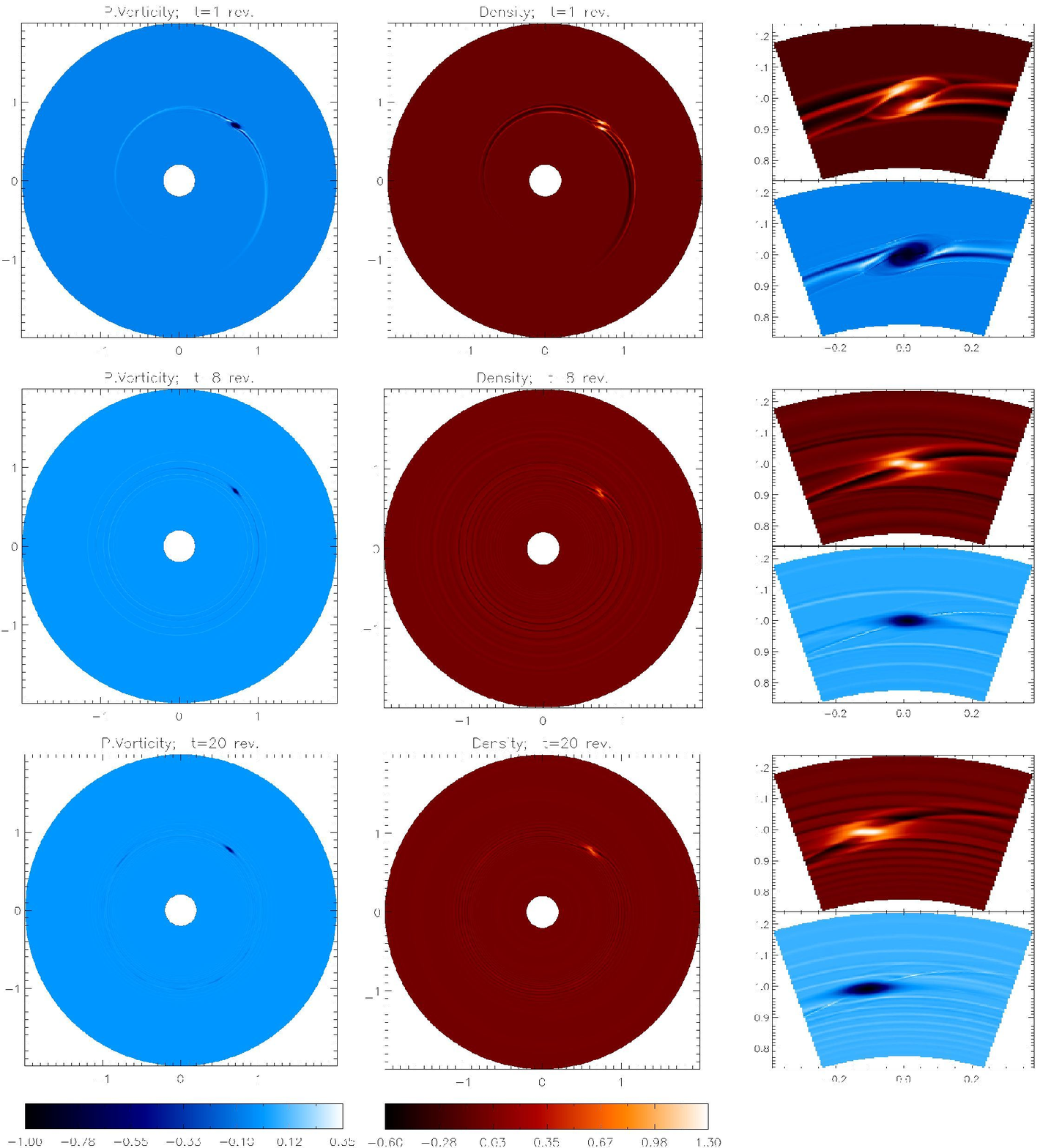}
 \caption {Dynamics of an initially imposed small-scale
 anticyclonic vortex (exponential distribution of vorticity, $q=1$, 
 $a=0.1$) with amplitude exceeding the second nonlinear
 threshold parameter: $\epsilon = 0.5$. The system parameters
 correspond to case C11 (see Table 2).  The first column shows the
 evolution of the perturbed potential vorticity and the second column 
 the evolution of density at times 1,8 and 20 revolutions (in units of
 the local disk revolution period). The initially imposed vortex with
 high amplitude emits the excess energy in form of density-spiral waves and
 within 4 revolutions adjusts to the single nonlinear
 configuration which survives further without significant variation
 throughout the entire period of simulation.
 See electronic version of the paper for color figures.
 } \label{Adjustment}
 \end{figure*}

Increasing the initial vortex amplitude, we found a second
nonlinear threshold:
\begin{equation}
 \epsilon^{**} = 0.25 \,.
\end{equation}
Vortices with $\epsilon>\epsilon^{**}$ experience direct adjustment
from the initial to the final  persistent structure, i.e. a
strong anticyclonic vortex. This process may be followed in Fig.
\ref{Adjustment}, where an initially unbalanced vortex, within $4$
revolutions, undergoes direct adjustment to the final nonlinear
configuration. The energy excess of the initial state is radiated
away in the form of spiral-density waves and shocks.

The final equilibrium configuration appears to be a nonlinear
attractor reached by the system if the initial amplitude exceeds
$\epsilon^{**}$ and  if the initial  spatial scale falls in a range
discussed in the next subsection. Indeed, the same nonlinear state
is developed from all initial vortices satisfying these conditions,
independently from the details of the initial potential vorticity
distribution (exponential or algebraic, circular or elliptic).

\subsection{Initial vortex size}
%
%

The result of nonlinear adjustment strongly depends also on the
initial vortex size. In order to quantify this dependence, we have
carried out computations with vortices of different initial size
$a$. The size of the vortex can be compared with the length-scale
\begin{equation}
 H \equiv c_s / \Omega_{\rm Kep} \,,
\end{equation}
which, in presence of the vertical component of gravity, describes the
scale height of a thin Keplerian disk. From a physical point of view,
a 2D analysis can give meaningful results only for vortices with sizes
larger than $H$. This will be our case of interest. For $c_s$ {\bf
  initially} constant all over the disk (as it is in our case), $H$ is
a function of the radial position and, at the vortex location, one has
$H = c_s$ (since $\Omega_{\rm Kep} (1) = 1$).

Our numerical results show that vortices undergoing direct nonlinear
adjustment have initial size in a range $a_{0} < a < a_{\max}$,
where the limits $a_{0}$ and $a_{\max}$ depend on the local disk
parameters. During the nonlinear adjustment the vortex decreases its
size, radiating the excess energy through spiral density waves and
shocks, and reaches the equilibrium configuration with size
$a_{0}$ that can be still larger than the disk scale height.
The final equilibrium configuration reached by these
vortices appears to be independent from the details of the initial
state and seems to represent a nonlinear attractor, whose
characteristics will be described in detail in the next Section. In
cases C1-C13 (i.e., $c_s=10^{-2}$), the value range of $a$ evolving
into the same nonlinear configuration is $2H < a < 10H$. For
comparison, in cases E2-E5 ($c_s = 10^{-3}$) this range is $12H < a
< 20H$.

\begin{figure*}[th!]
 \centering
 \includegraphics[width=\hsize]{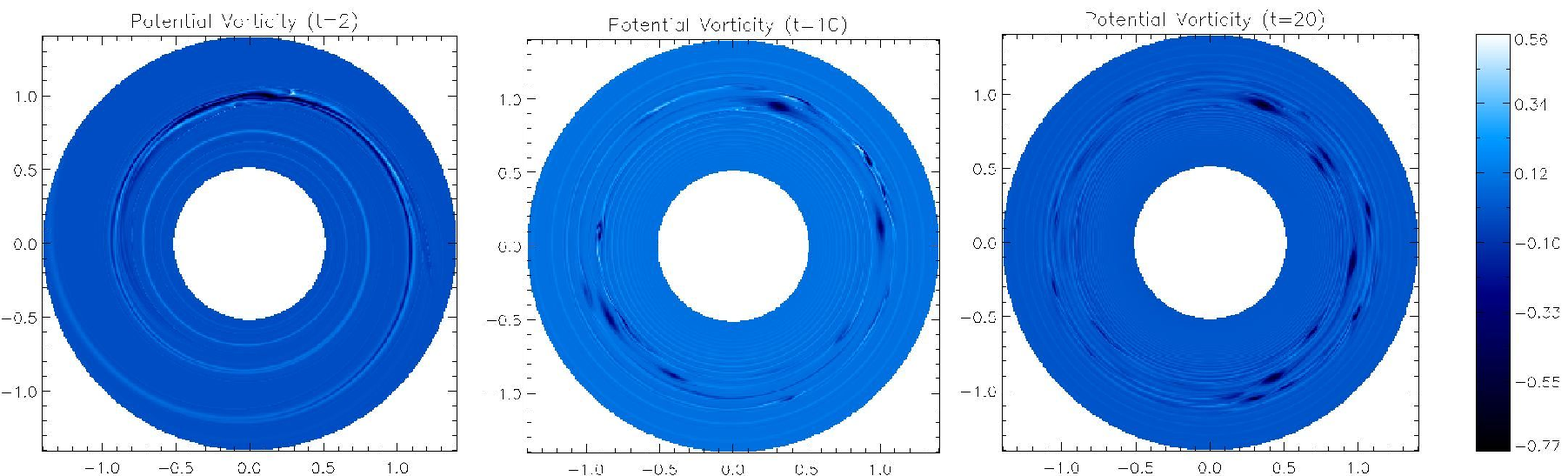}
 \caption{Potential vorticity evolution in the case of an anticyclonic
 vortex with exponential distribution of vorticity, $q=1$,
 $a=0.2$ and amplitude exceeding the second nonlinear threshold
 parameter: $\epsilon = 0.5$. The system parameters correspond to
 case C13 (see Table 2). The background Keplerian vorticity is
 subtracted. The dark spots reveal the anticyclonic patches. The
 radial transport of potential vorticity occurs on the initial
 stages, when the vortex is oversized and is subject to curvature
 radiation. At later times the background shearing stretches the
 vorticity and developed vortices only move azimuthally, while
 occasionally interacting and decaying. } \label{Oversize_PV}
 \end{figure*}

When the spatial-scale of the initial vortex exceeds $a_{\max}$, the
evolution is quite complex. We observe a radial transfer of
potential vorticity in both directions (see Fig \ref{Oversize_PV}),
caused by the combined action of shocks (radiated from the initially
imposed supersonic vortex) and of flow curvature (inducing Rossby
wave variations). The initial radial transfer of potential vorticity
is accompanied by a shearing process leading to a decrease of
potential vorticity localization and to a subsequent fragmentation.
The resulting picture tends to become even more complicated when
secondary vortices, formed at different radii and with different
sizes, start to interact. As a result, we see multiple vortices with
typical sizes equal to or smaller than $H$ leaning to further decay
in size and amplitude.

When the size of the vortex is smaller than $a_{0}$ and the initial
vortex amplitude exceeds the second nonlinear threshold, we observe
nonlinear adjustment to a final configuration with sizes
similar to the initial values. In this case, developed
vortices typically have sizes smaller than $a_{0}$.
However, if the initial amplitude largely exceeds the second
nonlinear threshold, we have found that the vortex can increase its
size and reach the nonlinear attractor discussed above. We have not
studied this case $(a < a_{0})$ in detail since the typical
size of vortices developed during the adjustment falls around $H$ or
smaller (unless the initial amplitude is quite high), indicating the
importance of 3D consideration.

\begin{figure}[th!]
 \centering
 \includegraphics[width=8cm]{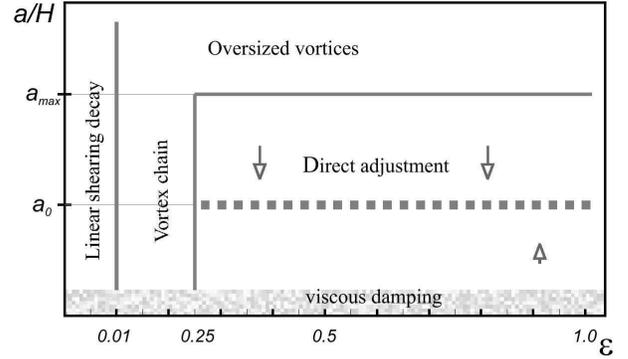}
 \caption{Sketch diagram illustrating the character of the evolution of
 an initially imposed anticyclonic vortex. In abscissa we have
 the amplitude ($\epsilon$) and in ordinate the size ($a$) of the initial
 anticyclonic vortex. Two nonlinear thresholds in amplitude
 (0.01 and 0.25) control the fate of the initial vortex, as it can
 decay linearly, evolve into a vortex chain or undergo direct
 adjustment to a single persistent configuration.
 The initial vortex size determines if the vortex will decay due to radial
 vorticity transport (oversized vortices), viscosity, or undergo
 direct adjustment. Horizontal dotted line shows the size of the
 equilibrium persistent configuration which can be interpreted
 as the nonlinear attractor.}
\label{fig:sketch}
\end{figure}

Fig. \ref{fig:sketch} shows a sketch diagram illustrating the
character of vortex evolution depending on its initial amplitude
($\epsilon$) and size ($a$).

In the following, we will focus our attention on the cases that
reach the equilibrium configuration. For them we discuss, in the
following subsection, the time-scales of nonlinear adjustment and,
in the following \S\ref{sec:structure}, the final equilibrium
structure.

\subsection{Time-scales of the adjustment}
%
%
%

\begin{figure*}[th!]
 \centering
 \includegraphics[width=\hsize]{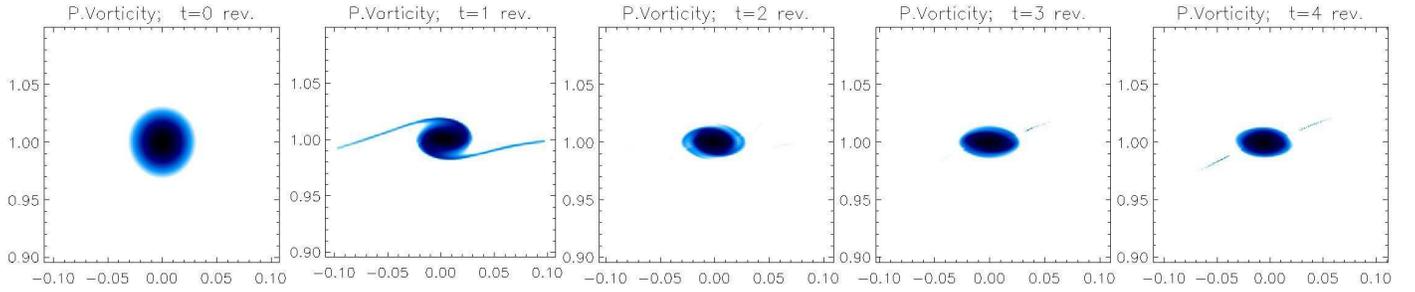}
 \caption{Evolution of  potential vorticity during the direct
 nonlinear adjustment. The initial conditions correspond to a small-scale
 anticyclonic vortex with $\epsilon = 0.5$, $a=0.05$ and $q=1$. The
 system parameters correspond to case C4 (see Table 2). Potential
 vorticity of the initially imposed vortical perturbations undergoes
 adjustment to the final equilibrium state within the first
 3 revolutions.} \label{VorticityAdjustment}
 \end{figure*}

The characteristic time-scale needed for radiating the energy excess
and reaching the equilibrium configuration strongly depends on the
vortex size. As we shall see in the next Section, the equilibrium
vortex configuration has a characteristic size that does not depend
on the initial size of the vortex. This means that bigger initial
vortices need to emit more energy to adjust to the final small-scale
configuration. Indeed, we have seen that a different number of
revolutions were needed to develop a stable long-lived structure
starting from vortices of different size:
 $$
 \begin{array}{lcl}
 a = 0.05 &\quad\rightarrow &\quad 3 \quad{\rm revolutions} , \\
 a = 0.1  &\quad\rightarrow &\quad 4 \quad{\rm revolutions} , \\
 a = 0.2  &\quad\rightarrow &\quad 6 \quad{\rm revolutions}
\footnote{Although multiple vortices
 are developed, here we measure the time for the development of the
 main vortex} . \\
 \end{array}
 $$
(runs C4,C11 and C13). The geometry of the potential
vorticity adjustment can be seen in Fig. \ref{VorticityAdjustment}.
It matches the patterns observed in previous numerical simulations
\citep[see e.g.][]{GodLiv99aa}. However, our high-resolution
simulations reveal new features of this process. The adjustments of
potential vorticity and density proceed in different ways and on
different timescales (see Fig. \ref{VorticityAdjustment}). When the
potential vorticity has almost reached its equilibrium
configuration, the density distribution still undergoes significant
structural variations. Initially it develops a {\it double core}
vortex configuration that slowly tends to change into a one core
stable vortex. For an initial value of $a=0.1$, the potential
vorticity adjusts in 4 revolution, while density needs 15
revolutions for reaching the final configuration \textbf{(case
C11)}. Moreover, the density adjustment (that is, the time needed
for the merging of the two density cores) shows a strong dependence
on the intrinsic numerical viscosity. Indeed, increasing the
resolution, this settlement time increases. The differences between
vorticity and density adjustments may be a consequence of the
existence of spiral shocks generated by the interaction of the
nonlinear vortex with the background shear flow, as we will discuss
in \S\ref{sec:waves}.

%

\section{Vortex stability and structure}\label{sec:structure}
%
%
%

\begin{figure*}[th!]
 \centering
 \includegraphics[width=\hsize]{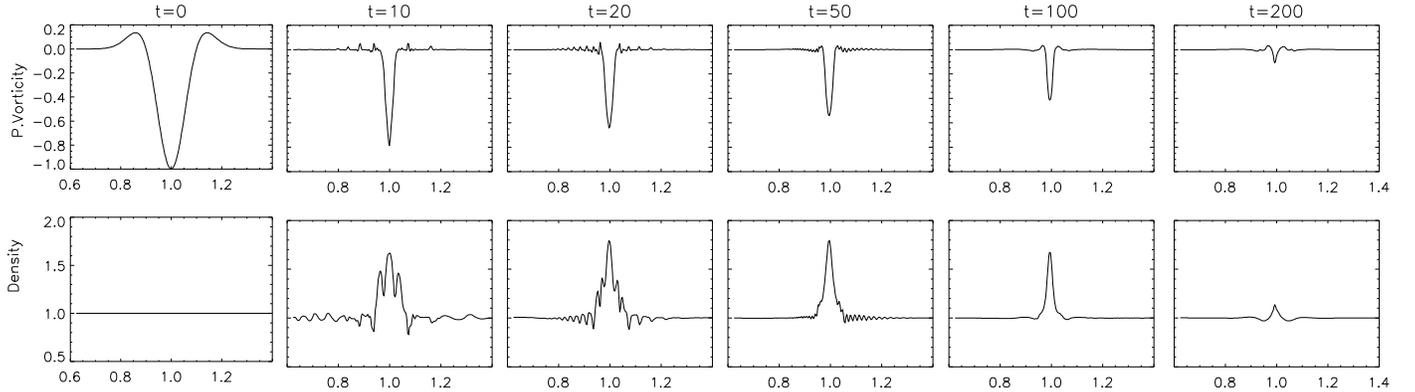}
 \caption{Evolution of  potential vorticity and  density for
 the  anticyclonic vortex with $\epsilon=0.5$, $a=0.1$,
 $q=1$ (setup C11). Radial cuts are taken across the
 geometrical center of the vortex (corresponding to a minimum in
 potential vorticity, for anticyclonic vortices). The figures show a
 gradual increase of density in the vortex core. Viscous forces
 damp oscillations around the vortex. At higher resolution this
 effect occurs on longer time-scales. After the initial adjustment,
 the vortex persists in the flow for long times and its dynamics is
 only a subject to viscous damping.} \label{2DStructure}
 \end{figure*}

One of the main goals of the present study is to describe the
stability and structure of long-lived  vortices in
Keplerian disks. For this purpose, we selected cases undergoing
direct adjustment to a single vortex, and we followed their long
time behavior. Fig \ref{2DStructure} shows radial profiles of
potential vorticity and density at the center of a vortex for case
C11, integrated for a total of $200$ revolutions. After 4-6
rotations the potential vorticity configuration shows a well defined
center. The density maximum matches the vorticity center after
approximately $15$ revolutions. We observe a profound persistent
structure throughout $200$ revolutions. Of particular interest is
the behavior of the mass accumulated by the anticyclonic velocity
circulation, shown in Fig. \ref{2DStructure}, where we observe a slow but
steady increase of density at the vortex center. The
same figure shows the steepening of the vorticity gradient during
the adjustment process.

\begin{figure}[th!]
 \centering
 \includegraphics[width=7.5cm]{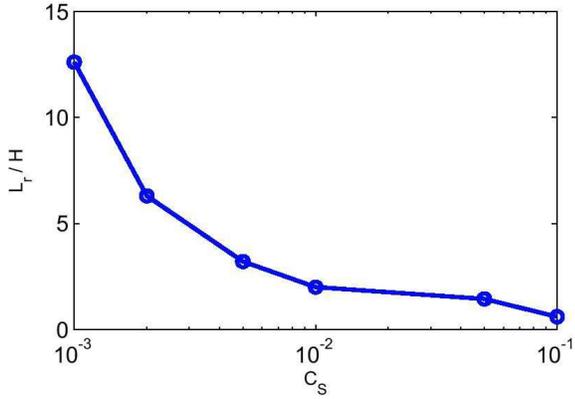}
 \caption{Scaling law for the  equilibrium vortex size:
 $f(c_{s})$ ($ = a_{0}/H$) as a function of sound speed
 $c_{s}$. The radial extent of the vortex is used.}
\label{fig:scaling}
\end{figure}

By analyzing vortices with different initial configurations, we have
found that the shape of the developed, persistent vortex depends
only on the disk sound speed $c_s$, and not on the characteristics
of the initial vortex (provided its size $a$ and amplitude
$\epsilon$ fall in the ranges discussed in the previous Section). We
can characterize the final state by its size $a_{0}$ (in the radial
direction) and ellipticity parameter $q$. The vortex size is
measured defining as vortex the region that has potential vorticity
larger than $10\%$ of the maximum value at the vortex center. The
following equation gives a scaling law for $a_{0}$ as function of
the sound speed $c_s$ \citep[see][]{BarMar05aa}:
\begin{equation}
 a_{0} = f(c_s) H \,,
\end{equation}
where the function $f(c_{s})$ (i.e. the ratio $a_{0}/H$) is shown in
Fig. \ref{fig:scaling}.
Note, that although our vortex scales with the sound speed (and the
disk height scale), we have found vortices with maximal size larger
than the ones described in \cite{BarMar05aa}. The plot shows that
the ratio $a_{0}/H$ increases for decreasing the sound speed, so the
condition for a 2D vortex behavior (vortex size larger than the disk
scale height), is verified to a better extent at lower sound speeds.
The final ellipticity $q$, on the other hand, appears to be almost
independent from $c_{s}$ and has a typical value $\sim 5$.

We confirm that the initial vortex configuration, for giving
direct adjustment, has to satisfy the condition derived in
\cite{DavSheCuz00aa}: the velocity gradient in the center of the
initial vortical  field must exceed the Keplerian gradient.
Moreover, we observe that the velocity gradient of the perturbation
tends to decrease during its evolution and reaches the Keplerian
value in the equilibrium configuration.

\begin{figure}[th!]
 \centering
 \includegraphics[width=8cm]{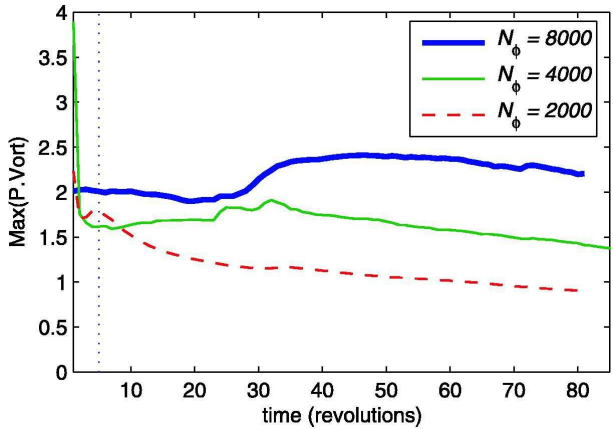}
 \caption{Plots of the maximum value of potential
 vorticity (st the vortex center) vs time, at three different
 resolutions. After the initial adjustment (vertical dotted line)
 vortex opposes viscous dissipation for some period and falls
 afterwards into the almost linear decay regime. At 2000 points in
 the azimuthal direction (12 grid points per vortex scale) vortex
 decays in about the 200 revolutions, while at 4000 point resolution
 the estimate of the vortex disappearance is at 350 revolutions. At
 8000 points the lifespan of the vortex is longer, but can not be confidently
 estimated using our data.}
 \label{Decay}
 \end{figure}

Fig. \ref{Decay} shows the decay curves of the maximum of potential
vorticity at the center of the vortex for different resolutions. As
expected, for finer meshes, the vortex decay is slower because of
the reduced numerical dissipation. However, apart from this
consideration, interesting effects can be noticed. After a rapid
variation within the adjustment time (first 5 revolutions in present
case), one can observe the formation of a stable persistent vortex.
Interestingly, the  vortex seems to be temporarily
able to oppose viscous dissipation, exhibiting for some time an
increase of the maximum potential vorticity. This effect implies
production of potential vorticity. On the other hand, potential
vorticity is a nonlinearly conserved quantity in barotropic flows.
Thus, the source of the coherent generation of potential vorticity
necessary to support or even enhance the vortex are the spiral shock
waves that will be discussed in \S\ref{sec:waves}.  They seem to
  be able to overcome the (numerical) dissipation at early times, but
  not at later times. The situation is quite complex, one of the
  possible reasons for this behavior could be a change in
  the internal vortex structure, like the transition from the double
  to the single density core.

\subsection{Cyclonic vortices}
%

The nonlinear dynamics of cyclonic vortices shows significant
differences from anticyclonic ones. In the limit of small
amplitudes, cyclonic vortices decay on the shearing timescale,
related to the geometrical stretching induced by the background
differential rotation. For higher amplitudes (exceeding the first
nonlinear threshold $\epsilon^*$) they exhibit a rapid decay on
time-scales shorter than the shearing time-scales. It then appears
that nonlinear forces accelerate the linear decay of cyclonic
vortices.

\section{Waves and shocks}\label{sec:waves}
%
%
%
%
%
%

Vortices in shear flows generate spiral density waves by a linear
mechanism first described by \cite{ChaTevBod97aa}, and further
investigated numerically in Keplerian flows by \cite{BodChaMur05aa}.
In the present context we study the characteristics of this process
when nonlinear forces are also in action. As clearly seen from Figs
\ref{Fragmentation}, \ref{Adjustment} and \ref{VorticityAdjustment},
the dynamics of potential vorticity is accompanied by wave emission,
not only during the transition time, but also afterwards, when the
nonlinear self-sustained vortex is fully developed. This enforce the
fact that nonlinearity does not suppress the wave emission.
Moreover, after the adjustment, waves emitted by the coherent vortex
structure appear to develop into spiral shocks.

\begin{figure*}[th!]
 \centering
 \includegraphics[width=17cm]{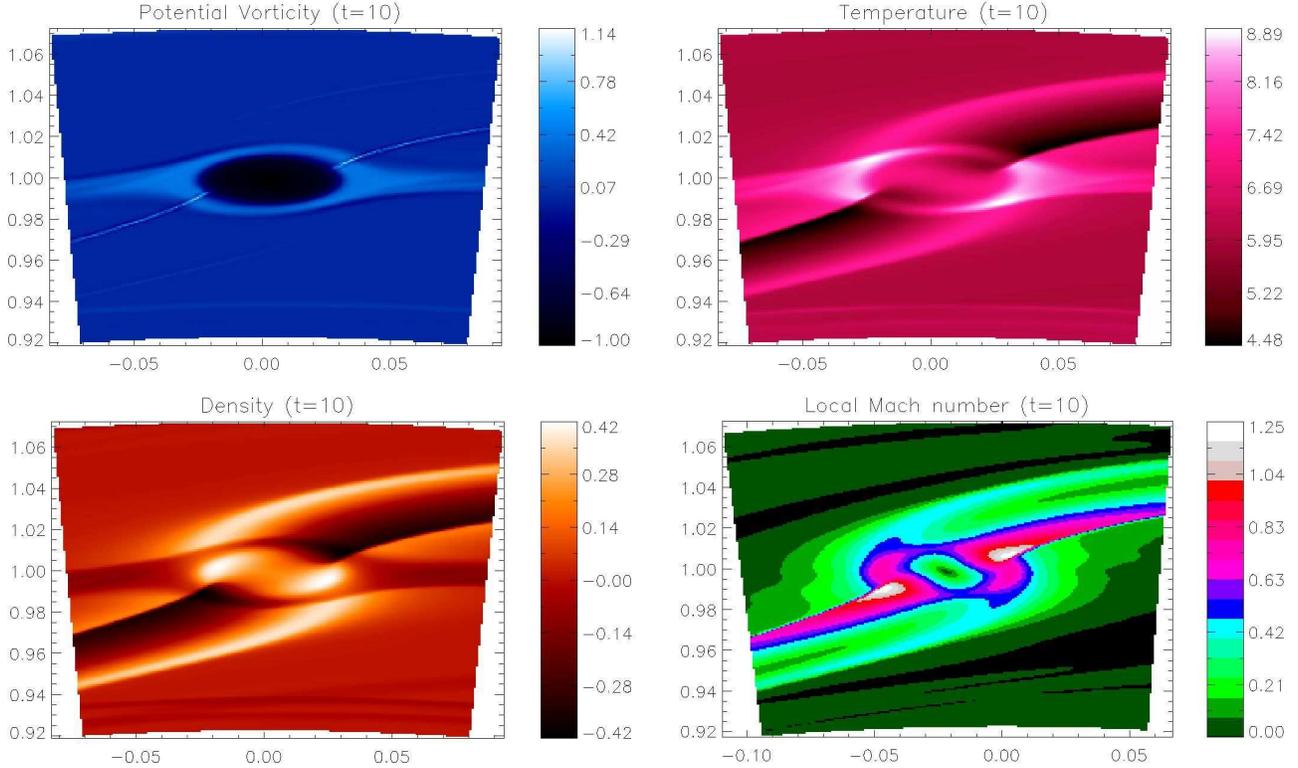}
 \caption{Potential vorticity, density, temperature and local
 Mach number of the anticyclonic vortex after 10 revolutions in the disk.
 Here the initial vortex has $\epsilon=0.5$, $a=0.05$, $q=1$ (case
 C4). Narrow spiral rays
 beaming out of the central vortex in the potential vorticity panel reveal
 spiral shocks attached to the outer radius of the vortex. Density
 perturbations show that compression induced by these shocks are
 responsible for the double core appearance of the vortex.
 The temperature plot shows the heating induced by  shock dissipation.
 Heating is stronger in the vicinity of the vortex leading to a
 consistent temperature increase of the global azimuthal ring that
 bears the  vortex. See electronic version of the
 paper for color figures. } \label{ShockSet}
 \end{figure*}

A regular structure of spiral shock waves has been found in
protoplanetary disk simulations with an embedded protoplanet
\citep{TanWat02aa, KolLiLin03aa}. Traces of shock generation by
vortices may be found in \cite{JohGam05aa}. In our simulations we
verify the existence of a steady pattern of spiral shocks produced
by a single  vortex. Moreover, shock waves appear to
be an inherent property of vortices in sheared compressible flows,
but they can be observed only at fairly high resolutions using shock
capturing schemes (limits on resolution will be discussed in
\S\ref{sec:numer}). Our high resolution calculations allow to study
these shock waves in great detail. In Fig \ref{ShockSet} we show the
distributions of potential vorticity, density, temperature and local
Mach number for the vortex and the attached shock waves. One can
distinctly recognize a wave-crest of the density-spiral wave
developing into a double shock configuration, with the shock ahead
(behind) of the vortex facing the outward (inward) region. A couple
of much weaker shocks, parallel to the strong ones, appear to be
present although they remain barely visible. These shocks strongly
affect the density structure of the developed vortex configuration,
resulting in a splitting of the vortex core. Eventually, however,
the shearing background leads to the merging of the cores, but the
shocks persist.

Spiral shocks induced by a wake of planets are believed, in some
situations, to be responsible for planet migration \citep[see][and
references therein]{PapTer06aa}. In our computations no radial
variation of the vortex position has been observed.

As seen in previous studies, spiral shocks affect dust accretion
rates on the vortex core and thus promote the formation of
planetesimal. In this sense, they increase the importance of
anticyclonic vortices in planetary formation.

The presence of shocks has consequences for the vortex evolution.
Nevertheless, the final fate of these structures cannot be easily
foreseen and requires much longer simulations. We can here only
sketch some possible scenarios. One possibility is the exhaustion of
matter in the vortex bearing ring and the formation of isolated
planetesimal. \cite{SchSpeGun04aa} have argued that spiral shocks
may lead to the gap formation. On the other hand, shocks heat the
ring at the radius where the vortex is sustained (see Fig.
\ref{ShockSet}), which in turn may trigger the linear Rossby wave
instability due to the unusual entropy gradient in the disk matter
\citep{LovLiCol99aa}. The instability will induce radial mixing and
possibly the destruction of the coherent vortices. On the other
hand, \cite{LubOgi98aa} have shown that spiral shocks can be
themselves unstable in three dimensions. Hence, as we said, longer
and possibly three-dimensional simulations can clarify this issue.

\section{Resolution study and numerical requirements}
\label{sec:numer}
%
%
%

Our direct numerical experience shows that a number of stringent
numerical requirements may be indispensable for observing the
correct dynamics of small-scale vortices in two-dimensional
compressible Keplerian flows. One necessary requirement is the need
to resolve small size perturbations.

\begin{figure}[th!]
 \centering
 \includegraphics[width=8cm]{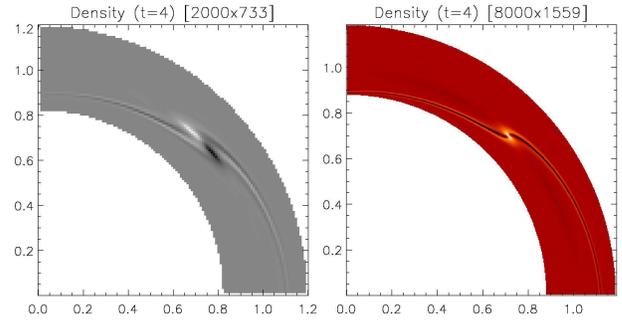}
 \caption{Dynamics of an initially imposed small-scale anticyclonic
 vortex with $q=1$, $a=0.01$ and $\epsilon=0.5$ at two different
 resolutions: ($2000\times733$, 6 grid points over the vortex) on the
 left and ($8000\times1559$, 24 grid points) on the right. The fate
 of the vortex adjustment strongly depends on the resolution. At
 lower resolution we see the development of a bipolar vortex. 
 Fully resolved vortex
 development needed resolution as high as 8000 points covering global
 azimuthal domain.} \label{ResolutionDefects}
 \end{figure}

Vortices resolved on $8$ computational zones are persistent and,
although affected by enhanced numerical viscosity and damping, still
survived more than 20 revolutions. Otherwise, at least 24 grid
points are necessary to correctly follow the nonlinear adjustment.
In Fig \ref{ResolutionDefects} we show an example of such a
situation, when numerical setup M (2000x733, 6 grid points over
vortex) failed to describe the direct nonlinear adjustment of the
initially imposed small-scale vortex, but displayed development of a
dipolar or even tripolar vortex configurations. Direct nonlinear
adjustment to the single long-lived structure is resolved only under
considerably higher resolution (setup VH, 8000x1559, 24 grid points
over the vortex) thus setting overall restriction on the resolution
of our code.

 Notwithstanding the fact that the purpose of the simulations is the
 study of small-scale vortices,  global simulations
 are needed: vortices interact globally and change the density,
 temperature and potential vorticity in the whole ring at the radius
 where they are situated. Hence, to get a correct dynamical picture, we
 need true global simulations with a domain covering $2\pi$ in azimuthal
 direction. Calculations on smaller domains in $\phi$ ($\pi/2$) have
 lead to drastically different results, since the two vortices at
 the same radius start to interact before the nonlinear adjustment
 occurs.

Another requirement on the numerical code is that it should be able
to capture shocks, an important ingredient of the overall balance of
the sustained structures in differentially rotating flows. In our
case the PLUTO code employs solvers that accurately resolve the shock
dynamics. This requirement does not play in favor of spectral codes
for compressible simulations of the Keplerian disk dynamics.

\section{Summary}\label{sec:summary}
%
%
%
%

Direct numerical simulations of the nonlinear dynamics of
coherent vortices in 2D compressible protoplanetary disks
with Keplerian differential rotation have been presented.

We have shown the possible existence of anticyclonic vortices with
sizes exceeding the Keplerian disk height scales. We have followed
the evolution of such vortices for 200 local revolutions, showing
their persistence and stability.

We have found that the development of long-lived 
nonlinear anticyclonic vortex configurations occurs only when the
amplitudes of the initially imposed vortex perturbations exceed some
threshold value. We have distinguished two nonlinear threshold
values, that define the fate of the initial vortex. At small
amplitudes, the vortex behaves linearly and it decays due to the
shearing deformation. When the amplitude exceeds the first
threshold, the vortex stretching is followed by nonlinear
instabilities, leading to the formation of multiple small-scale
anticyclonic vortices. At amplitudes higher then the second
threshold, direct transition from the initially unbalanced to the
final equilibrium configuration occurs. We have interpreted the
latter process as the nonlinear vortex adjustment and studied the
parameters that can describe this process. We found that the vortex
adjustment proceeds on two different time-scales. Shorter times are
required for the settling the potential vorticity (3-6 revolutions)
and longer time-scales are required for the adjustment of density.
Density adjustment time-scales depend on the amount of viscosity
present in the flow. Thus, we observe a transient double-core
vortex, that evolves into a single core structure. The density core
splitting of the  vortex is due to the spiral shock
induced compression. Cores tend to merge at later times and the
process is promoted by the intrinsic numerical viscosity. At lower
viscosity, double core vortices survive for longer times. The
density at the center of the vortex slowly increases and the
potential vorticity gradient steepens.

The structure of the developed long-lived vortex does not depends on
the initial vortex configuration, provided it exceeds the second
threshold amplitude and its size does not exceed a limiting value.
In this sense we found a nonlinear attractor that is the final
configuration of a wide range of initial vortical perturbations. We
found a scaling law for the equilibrium vortex size. In
particular we find that the ratio of the vortex size to the local
disk scale height increases with the decrease of the sound speed.
Therefore, at low sound speed the radial extent of  equilibrium anticyclonic
vortices can significantly exceed the disk thickness. 
In this case vortex dynamics is intrinsically 2D and should
be well modelled by the present simulations. Note that the sound
  speed required is somewhat lower than the typical values expected for
  protoplanetary disks $(c_{s} \sim 0.05 - 0.1)$. In the expected range the
  radial  extent would be of the order of the scale height, while the
  azimuthal extent will be five times larger, in this case, to have a
  conclusive answer,  3D calculations are needed.  

Vortices generate density-spiral waves that rapidly develop into
shocks. As a result, a long-lived nonlinearly balanced vortex is
accompanied by two spiral compressible shock waves facing  both
radial directions. The shock strength is maximal in the vicinity of
the vortex outer edge, where the shock is attached to the main
vortex circulation and heats all the annular disk region containing
the vortex. Higher resolution runs indicate that these shock
waves are able to generate local potential vorticity and may support the
vortex against viscous dissipation.

We analyzed cyclonic vortices at nonlinear amplitudes. It seems that
the linear decay due to the shearing deformation is accelerated by
nonlinear effects.

We performed a resolution study and found the necessity of a global
high-resolution approach to numerical simulations of these
processes. The fate of the vortex depends significantly on the
dissipation properties of the code. So, in this sense the use of
physical viscosity, as opposed to the intrinsic numerical viscosity
present in our simulations, would improve understanding of the
vortex behavior and damping.

Our study contributes to the scenario of planetary formation inside
the core of the long-lived vortices. We found that protoplanetary
disks with lower sound speed can sustain vortices with higher ratio
of vortex size to disk thickness and create a more favorable
conditions for dust trapping and mass accumulation. In this context,
we have also found a steady increase of density inside the
nonlinearly balanced vortex, partly, due to the existence of
persistent, steady, spiral shock waves that we showed to be  an
intrinsic property of stationary vortices with sizes exceeding the half
thickness of the disk.

\begin{acknowledgements}

This work is supported by ISTC grant  G-1217 and by italian MIUR.
 A.G.T. would like to acknowledge the hospitality
of Osservatorio Asrtonomico di Torino. Numerical calculations have
been partly performed in CINECA (Bologna, Italy) thanks to the
support by INAF.

\end{acknowledgements}


\clearpage

\end{document}